\documentstyle{mn2e}

\input{psfig}
\newcommand{\be}{\begin{equation}}
\newcommand{\ba}{\begin{eqnarray}}
\newcommand{\ee}{\end{equation}}
\newcommand{\ea}{\end{eqnarray}}
\newcommand{\etal}{et al.\ }
\def\lesssim{\mathrel{\hbox{\rlap{\hbox{\lower4pt\hbox{$\sim$}}}\hbox{$<$}}}}
\def\gtrsim{\mathrel{\hbox{\rlap{\hbox{\lower4pt\hbox{$\sim$}}}\hbox{$>$}}}}
\def\simless{\mathbin{\lower 3pt\hbox
  {$\rlap{\raise 5pt\hbox{$\char'074$}}\mathchar''7218$}}}  % < or of order
\def\simgreat{\mathbin{\lower 3pt\hbox
  {$\rlap{\raise 5pt\hbox{$\char'076$}}\mathchar''7218$}}}  % > or of order

\def\apj{ApJ}

\def\apjl{ApJL}
\def\aap{A\&A}

\def\mnras{MNRAS}

\begin{document}

%%%%%%%%%%%%%%%%%%%%%%%%%%%%%%%%%%%%%%%%%%%%%%%%%%%%%%%%%%%%%%%%%%%%%%%%%%

\title[Chaos and Particle Orbits in Dark  Haloes]
      {Influence of Orbital Behaviour and Chaos on the Shape of Dark Matter Halos}

\author[A.V. Macci\`o et al.]          
       {\parbox[t]{\textwidth}{
        Andrea V. Macci\`o$^{1,2}$\thanks{E-mail:maccio@mpia.de},  
         Ioannis Sideris$^2$, Marco Miranda$^{2}$, Ben Moore$^{2}$ \\
        Roland Jesseit$^{3}$ }  
        \vspace*{3pt} \\
       $^1$Max-Planck-Institute for Astronomy, K\"onigstuhl 17,  
           D-69117 Heidelberg, Germany\\ 
       $^2$Institute for  Theoretical  Physics, University  of
           Zurich,  CH-8057, Zurich,  Switzerland\\  
       $^3$Universit\"ats Sternwarte M\"unchen, Scheinerstrasse 1, D-81679 M\"unchen, Germany }

%%%%%%%%%%%%%%%%%%%%%%%%%%%%%%%%%%%%%%%%%%%%%%%%%%%%%%%%%%%%%%%%%%%%%%%%%%

\date{}

\pubyear{2007}

\maketitle

\label{firstpage}

%%%%%%%%%%%%%%%%%%%%%%%%%%%%%%%%%%%%%%%%%%%%%%%%%%%%%%%%%%%%%%%%%%%%%%%%%%

\begin{abstract}

It has been  shown that  the dissipative  gas infall
during galaxy formation has the capability to modify the shape of dark
matter halos.  In this paper we perform the first detailed analysis of
particle orbits in a cosmological dark matter halos to understand
{\it how} and {\it why} baryons alter its shape. We perform a series of
numerical experiments where we grow  a baryonic core inside a live dark
matter halo  extracted from a  cosmological simulation. We  follow the
orbits of more than 300  particles with more that 50000 timesteps. Our
results clearly show that the dissipational  component is responsible
for repeatedly deflecting  orbits which visit often the  center of the
system. Moreover the gravitational potential time dependence associated with 
the growth of the baryonic mass, shifts the frequencies of the orbits, making
them extremely chaotic. This randomization makes the orbits explore a 
large phase space. When this effect takes place for a significant
number of orbits it will be manifested in the density distribution as
an approach to a rounder configuration. As a consequence, 
the influence of the central mass on the shape of the phase space 
decreases with increasing distance from the center. 
We discuss the importance of future analysis of controlled experiments
(i.e. using analytic potentials instead of live DM halos) 
to better decipher the dynamics of this phenomenon.
\end{abstract}

%%%%%%%%%%%%%%%%%%%%%%%%%%%%%%%%%%%%%%%%%%%%%%%%%%%%%%%%%%%%%%%%%%%%%%%%%%


%%%%%%%%%%%%%%%%%%%%%%%%%%%%%%%%%%%%%%%%%%%%%%%%%%%%%%%%%%%%%%%%%%%%%%%%%%

\section{Introduction}
\label{sec:intro}

Currently, the favored cold dark matter (CDM) model of structure formation suggests that dark matter (DM) halos of galaxies and clusters should admit triaxial density profile (Barnes \& Efstathiou 1987, Dubinski \& Carlberg 1991, Warren et al. 1992, Jing \& Suto  2002, Moore et al. 2004). This expectation is well-accepted, but (in principle) it could also be tested against observations. In this respect, triaxial dark halos surrounding both elliptical and disk galaxies could, not only provide clues about the underlying physics of galaxy formation, but also impose constraints on different formation scenarios (Dubinsky \& Carlberg 1991). On the other hand, the recently observed coherence of the tidal stream of the Sagittarius dwarf spheroidal suggests a nearly spherical halo configuration for the Milky Way (Ibata et al. 2001, Mayer et al. 2002, Majewski et al. 2004), in contradiction with what has been obtained by cosmological simulations (i.e. Jing \& Suto 2002, Allgood et. al 2005, Macci\`o et al 2007).

This inconsistency could possibly be resolved by carefully examining the effect gas cooling has on halo shapes. In 1994 Dubinsky studied the effects of dissipative gas infall on the shape of DM halos by adiabatically growing central-mass concentrations inside initially triaxial DM halos. A more extensive study was made recently by Kazantzidis et al. (2004; K04 hereafter), where a number of cosmological gas dynamics simulations were performed; the main result was that halos formed in simulations with gas cooling are significantly more spherical than corresponding halos formed in adiabatic simulations.

Aiming to extend the aforementioned investigations (Dubinsky 1994 and K04), 
this paper attempts to address the question of {\it why} and {\it how} the gas 
dissipation alters the halos shape. The main addition is that we approach the 
problem from a dynamical point of view. 

The relation between particle orbital properties and the shape of bound structures 
has been already studied by several authors (i.e. Jesseit, Naab \& Burkert 2005, 
Naab, Jesseit \& Burkert 2006, Valluri et al 2007).  
In this paper we present the first broad analysis of particle orbits in cosmological 
DM halos, and we 
investigate how these orbits modify their behaviours when a baryonic core is grown 
inside the original triaxial dark matter halo.

The analysis of the orbits is two-leveled. Firstly, we study them in a statistical fashion, investigating quantities like the evolution of axes ratios of the configuration. Secondly, we examine the behaviour of individual orbits; we ask how their shapes are affected when there is a dissipational component at the center of the configuration, but we also  employ a  frequency  analysis scheme  to quantify  the chaoticities, and we attempt to investigate how chaos connects to the change of their shapes.

This paper is organized as follows: in section \ref{sec:numsim} we present our codes and modeling; results are presented in section \ref{sec:numexper}; section \ref{sec:discussion} contains our conclusions and description of follow-up for this study.

\section{Numerical Simulations}
\label{sec:numsim}

All simulations were performed  using PKDGRAV, a tree-code for cosmological simulations of $N$ bodies, written by J. Stadel and T. Quinn (Stadel 2001). This code uses spline kernel softening, for which the forces become completely newtonian at two softening lengths. Individual time-steps for each particle are chosen to be proportional to the square root of the softening length $\epsilon$, over the acceleration, $a$: $\Delta t_i=\eta\sqrt{\epsilon/a_i}$. Forces are computed using terms up to hexadecapole order and a node-opening angle $\theta$ which we set to $0.6$.

The original dark matter halo was extracted from an existing $N$-body simulation where we had followed the evolution of 250$^3$ particles in a volume of 20 Mpc with a softening $\epsilon=0.6$kpc. We adopted a flat $\Lambda$CDM cosmology with parameters from the first year WMAP results (Spergel \etal 2003): matter density $\Omega_M=0.268$, baryon density $\Omega_b=0.044$,  Hubble  constant  $h \equiv  H_0/(100  \rm km\,s^{-1}Mpc^{-1})=0.71$, and  a scale-invariant, Harrison-Zeldovich power-spectrum with normalization $\sigma_8=0.9$. 
A detailed analysis of this simulation has been already presented 
in Macci\`o et al. (2007).

At redshift zero we selected a galaxy-like halo having the following characteristics: M$_{vir}= 1.7  \times 10^{12}  \rm M_{\odot}$, R$_{vir}=248$ kpc  and about 110000 particles within the virial radius. This halo is well-fitted by an NFW (Navarro, Frenk and White 1997) profile with a concentration parameter of 9.5. We also computed the shape of this halo with the same technique described in Macci\`o et al. (2007), getting for the axis ratio $b/a=0.82$ and $c/a=0.71$, where $a,b,c$ are the long intermediate and short axis respectively. For our further studies we cut from the original simulation a cubic region of 1x1x1 Mpc h$^{-1}$ centered around the selected halo. This region sets the new simulation volume for the baryonic mass growth experiment.

\subsection{Baryonic core modeling}
\label{sub:BarCore}

The investigation of galaxy formation inside dark matter halos is a challenging problem. In hierarchical models a collection of smaller scale fluctuations (protogalaxies) merge into forming a smooth, centrally concentrated galaxy. Both gravitational collapse and dissipation occur at the same time; if the cooling time of dissipation is shorter than the gravitational free-fall time, then gas will sink into the center of the system to form a luminous galaxy. Following Dubinski (1994) we modeled the formation  of the baryonic core of our galaxy-model by slowly growing a mass distribution at the center of a previously selected dark matter halo. The mass-growth follows a linear law, until it reaches a maximum mass comparable to the expected luminous mass of galaxies. We chose $M_b=1.8 \times 10^{10} \rm M_{\odot}$ for the final baryonic mass and $T=7$ Gyrs as growth time (this is the time elapsed from $z=1$ to $z=0$ in our $\Lambda$CDM universe). We also performed two more simulations with a smaller final mass for the baryonic core (keeping constant the growth time), namely we used a $1/4$ and $1/16$ of the previous value for $M_b$. For comparison purposes, we also performed a simulation with no growing mass at the center, where we simply allowed our halo to evolve in isolation (i.e. once extracted from the $N$-body simulation) for 7 Gyrs. After that time we stopped the mass growth on the baryonic core, and we let all the aforementioned simulations evolve for an additional 7 Gyrs, in order for our halos to reach a well-relaxed configuration.

In first approximation, we treated the luminous component of the galaxy as a single particle within the simulation. At time $T=0$ we gave to this new particle the same coordinates and velocities components of the dark halo center of mass. This particle was then free to move in the usual self-consistent fashion of an $N$-body simulation, but, in practice, it quickly settled at the center of mass of the dark matter halo because of dynamical friction.

One may object that this zeroth-order approximation might generate unphysical effects; for example having only one particle at the center of the configuration may cause unphysical close encounters with the dark matter particles, with subsequent scattering of the last. To assure that this was not the case in our experiments, we quantified the importance of these effects in order to separate physical phenomena from numerical artifacts generated by the approximations of our schemes.

To this end we performed two additional simulations. Keeping constant the growth-rate of the central mass we: i) increased the kernel softening for the growing particle to four times the one on the surrounding $dm$ particles, from 0.6 kpc to 2.4 kpc, and ii) instead of using one particle we split its mass into 500 particles. The initial spacial size of this particle-cluster was 5 kpc and all the particles increased their masses following the same growth-rate. As is clearly stated in the discussion section our results suggest that there is no significant, quantitative or qualitative, variation between these two simulations and the original one (one particle and small softening). This encourages us to be confident about the physical (as opposed to the numerical) origin of our results.

\section{Numerical Experiments}
\label{sec:numexper}

We performed a number of simulations of collapsing halos with and without a growing 
central baryonic mass. The density configuration of the initial conditions was 
triaxial with axes $a>b>c$. In these simulations we selected the same sample of 300 
particles to analyze. For these particles we recorded their positions and velocities 
for more than 55.000 time-steps throughout the total evolution time (14 Gyrs). 
The recording time-step of the orbits was set equal to the smallest time-step 
achieved by each particle during the force integration. Those {\it tracked} particles 
uniformly cover the energy distribution of all the particles at $T=0$ and their 
distance from the center ranges from 4 to 150 kpc. We also tracked the orbits of 
all the growing particles although we did not use them for the orbits modification 
analysis.

We investigated carefully both the behaviours of individual orbits, and the way these 
behaviours affect the morphological nature of the central regions of our systems. 
The evolutions of the axes ratios of both dissipational and dissipationless regimes 
were determined by employing an analysis of their inertial tensors.

\begin{figure*}
\psfig{figure=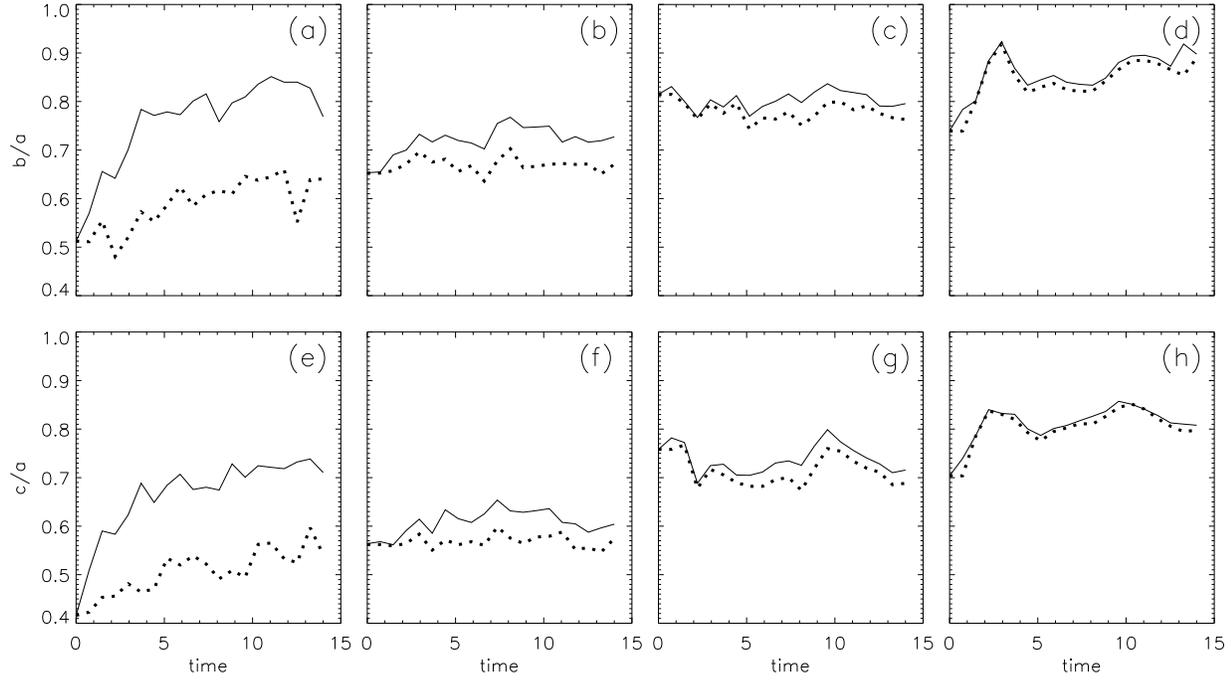,width=500pt}

\caption{\label{fig:fig1.ps}
Top row panels: evolution of the axes ratio $b/a$ of two systems, one without central baryonic mass (dotted line) and one with central baryonic mass (solid line) for different distances $r$ from the center. 
(a) $r=14.9kpc$, 
(b) $r=163.4kpc$, 
(c) $r=311.9kpc$, 
(d) $r=460.4kpc$. 
Bottom row panels: same as top row panels but for axes ratio $c/a$. 
(e) $r=14.9kpc$, 
(f) $r=163.4kpc$, 
(g) $r=311.9kpc$, 
(h) $r=460.4kpc$.}
\end{figure*}

\begin{figure*}
\psfig{figure=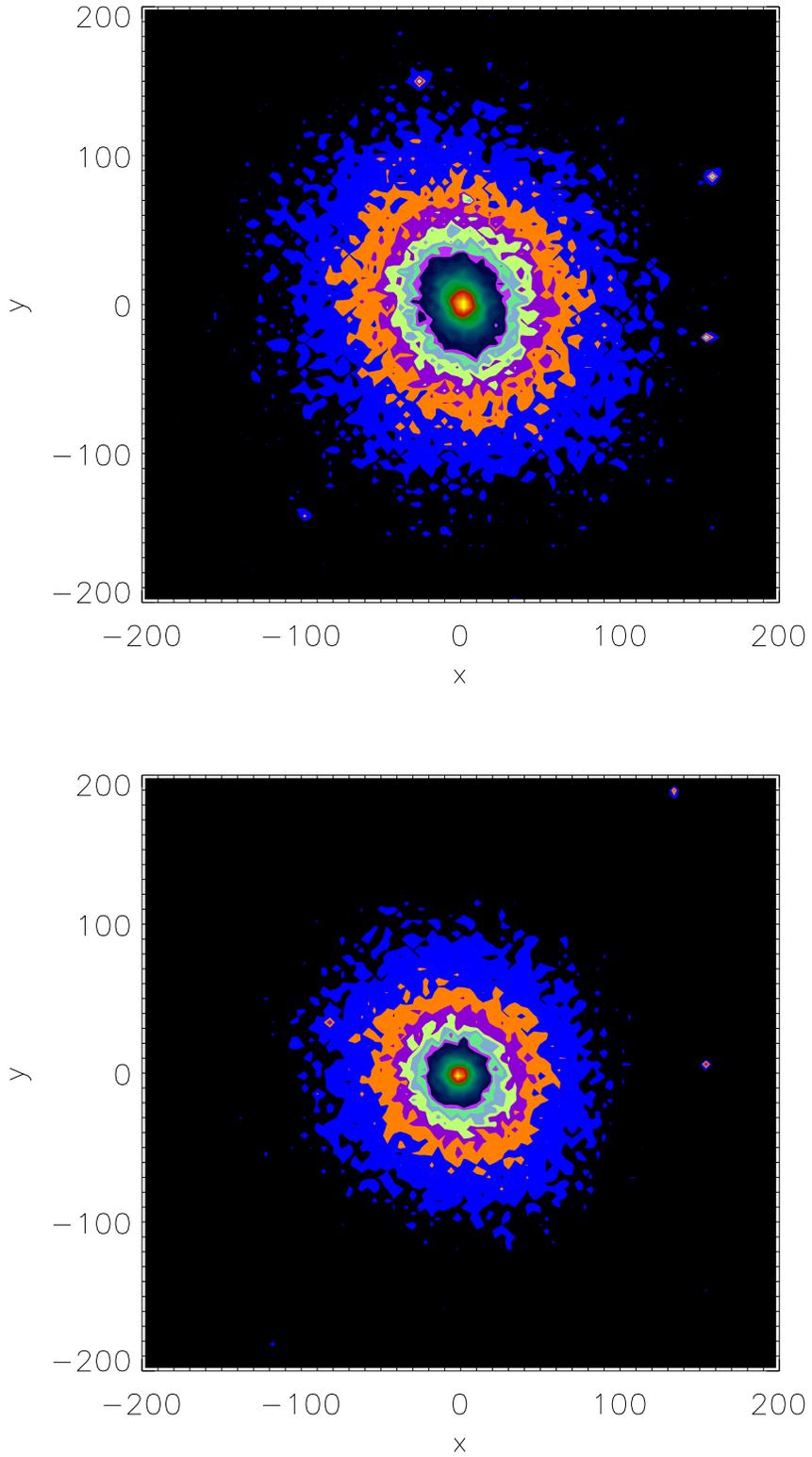,width=350pt}

\caption{\label{fig:fig2.ps}
Density contours for the central parts of a dissipationless (top) and dissipational (bottom) halo at the end of the evolution (14Gy). There is an obvious rounding of the bottom configuration.}

\end{figure*}

The general phenomenology seen already in previous works (K04), emerged in our 
simulations too; in the absence of a central baryonic mass, the original triaxiality 
insists throughout the evolution of the system. On the other hand, when the central 
dissipational component is present, there is a rapid increase of the values of the 
axes ratios $b/a$ and $c/a$ which is more evident at the central parts of the system,
and intensifies as time evolves (Fig.~\ref{fig:fig1.ps}). This increase is manifested
as a visual ``rounding''  of  the  contours of the density configuration 
(Fig.~\ref{fig:fig2.ps}).

Several investigators (Norman et al., 1985, Gerhard and Binney, 1985, Barnes and Hernquist, 1996,Merritt and Quinlan, 1998, Valluri, and Merritt, 1998) have provided a dynamical explanation for a similar behaviour occurring in the context of galactic systems with central black holes. They have shown that black holes alter the potential at the center of the system to a rounder one, and they modify several of the box orbits and tube orbits with low angular momentum (these orbits naturally visit the center of the system and serve as important building  blocks  for the  triaxial backbone of ellipticals.) Also, the phase space at the center should follow the potential and become rounder too. Since a significant number of orbits behave chaotically they will attempt to cover all the phase space energetically available to them. Therefore one expects the motion of orbits to build density configurations with rounder shapes.

A comment about timescales of chaotic diffusion may be useful here. In general, chaotic orbits attempt to access all the phase space energetically available to them.  However, the timescale necessary to achieve that, relates closely to the dynamical structure of the phase space. When parts of the phase space are occupied by regular orbits, chaotic orbits located very close to them try to behave like regular too, and remain constrained in restricted regions of the phase space for a long time, often longer than the age of the system (Contopoulos 2002, Efthymiopoulos et al. 1997, Sideris 2006); these orbits are usually called ``sticky'' or ``weakly chaotic''. Sticky orbits may need a significantly long time to escape into a wider chaotic sea and thus become ``wildly chaotic''. Only then they can occupy a much broader phase space area. On top of that, in a three-dimensional phase space although different chaotic areas are interconnected, it is the location and extent of the regular regions which determines how fast the diffusion of chaotic orbits occurs, and whether different parts of a chaotic sea can work as ``bottlenecks'' of the evolution.

Another issue is the influence of the time-dependence. In the experiments we performed there was the significant time-dependence associated with the growth of the central mass. In time-dependent regimes every energy at every time is associated with its own phase space structure. Orbits do not conserve energy, they move from energy level to energy level, thus they visit a plethora of different phase spaces; this means that their nature can actually alternate from regular to chaotic and viceversa (Kandrup et al. 2003).

We investigated the chaoticity of a sample of 300 orbits randomly chosen from the orbits located at the central part of our density configuration. The maximum initial radius of the sample was about 150 kpc. The first question we asked was: how does the chaoticity of the dissipational system compare to the chaoticity of the dissipationless one?

Regular orbits are associated with a limited number of frequencies of motion (these being the main frequency per degree of freedom and its harmonics). Theoretically these  frequencies should be singular, e.g. delta-functions. On the other hand, chaotic orbits are associated with many frequencies; their spectra are broad and theoretically continuous (Lichtenberg \& Lieberman 1992).

Numerically one does not have the luxury to integrate orbits for infinite or even sufficiently long time. This causes some unavoidable numerical artifacts; frequencies of regular orbits, although very localized, appear to have small but measurable tails in the power spectrum.  Frequencies  of  chaotic  orbits  are  numerous  but discrete. Despite these difficulties, one can usually distinguish easily between regular and chaotic orbits, provided that the orbit has been integrated for an adequate number of orbital periods. As the number of available orbital periods decreases, our confidence regarding the characterization of the orbit decreases accordingly. A method of computing localized  frequencies in localized  time would  be invaluable, especially for limited-time evolutions, but such method has not been clearly formulated yet.

In this spirit, we quantified chaos using a straightforward numerical technique associated with Fourier analysis ( Press et al. 1993; Kandrup et al. 1997). For every orbit we computed its three Fourier spectra, one per degree of freedom. Then (after we sorted the frequencies in descending order of their power) we simply counted how many of the emerging frequencies consist the 90\% of the power spectrum (starting with the strongest frequency and moving down to weaker ones). (Similar experiments with 80\% and 95\% gave similar results qualitatively). Finally we added the numbers of all three dimensions:

\begin{equation}
n_{0.90}=n_{x 0.90}+n_{y 0.90}+n_{z 0.90}
\end{equation}

\begin{figure*}
\psfig{figure=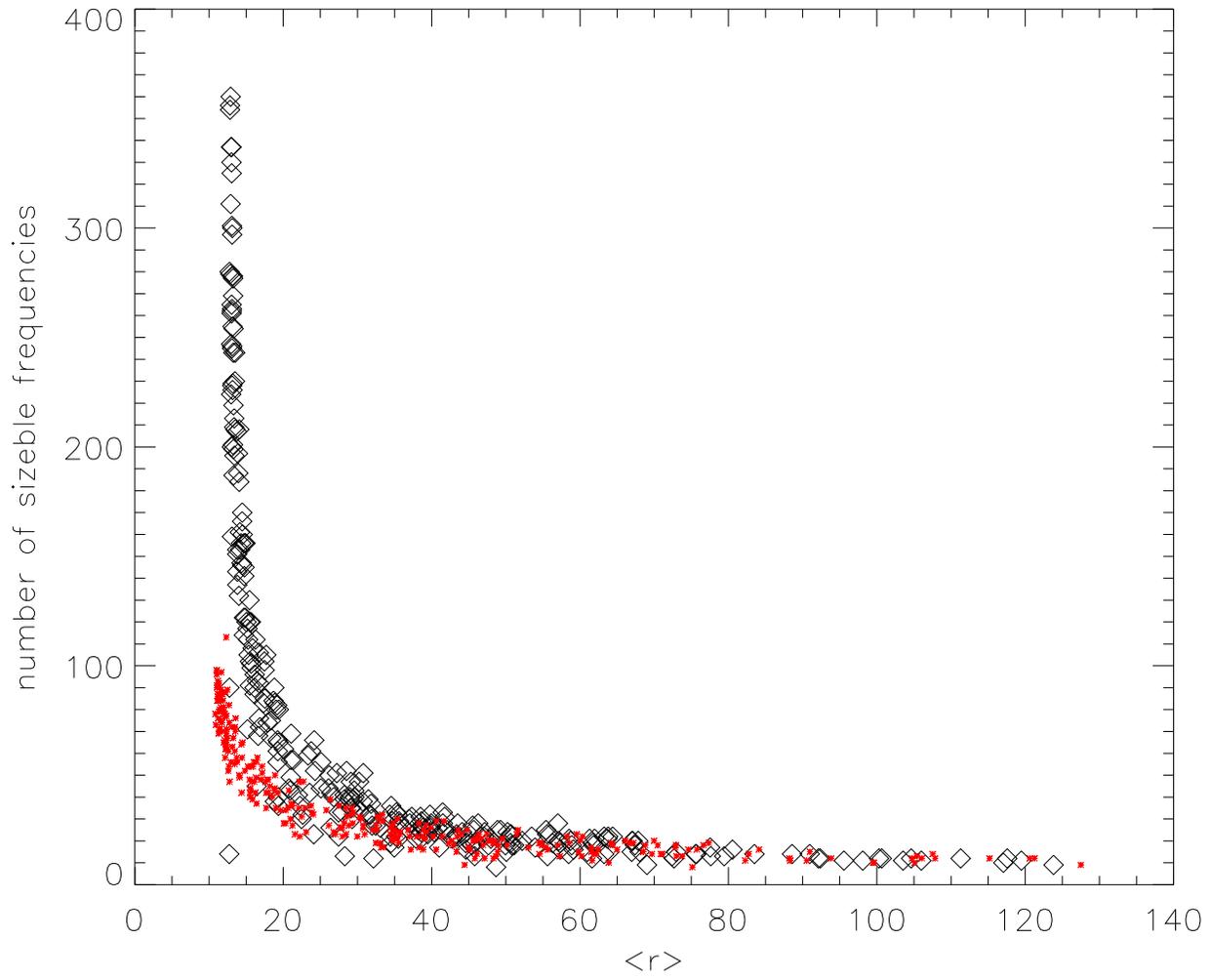,width=500pt}
\caption{\label{fig:fig3.ps}
Complexities (number of important frequencies of motion) for 300 orbits versus mean radial distance from the center. Red asterisks represent the dissipationless points while black diamonds the dissipational ones.}
\end{figure*}

The results of this analysis suggest that, when there is dissipation, chaos increases significantly at least close to the center of the system (Fig.~\ref{fig:fig3.ps}).  The characterization results for orbits moving close to the center are reliable since the number of orbital periods is about 30-50. As one moves away from the center, the number of orbital periods decreases and the reliability of the characterization unavoidably deteriorates. Still the trend close to the center is obvious; the dissipational component makes the system more chaotic.

\begin{figure*}
\psfig{figure=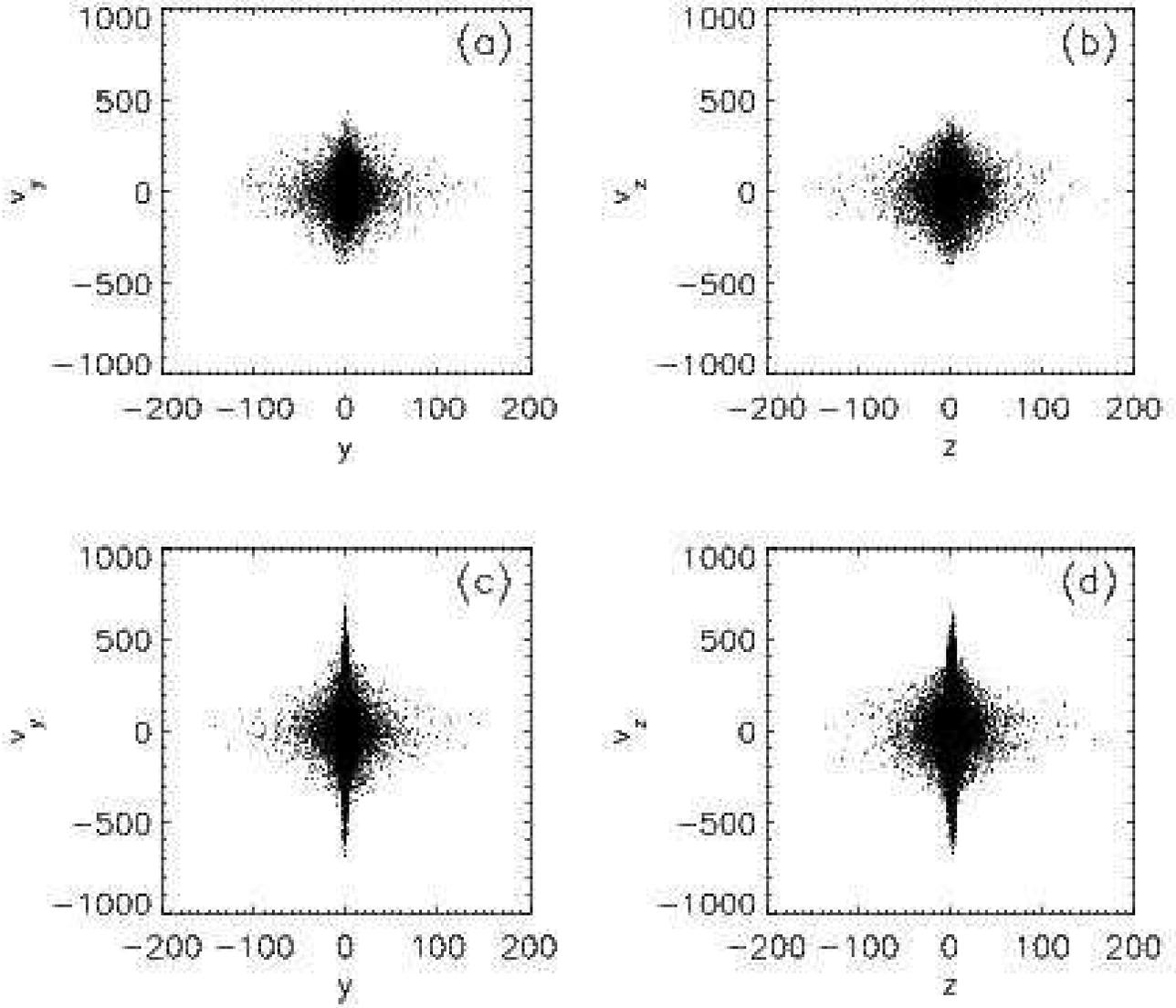,width=500pt}
\caption{\label{fig:fig4.ps}
(a) and (b) Poincar\'e sections of the dissipational points recorded at $v_x=0$. (c) and (d) same for the  simulation with a growing central baryonic mass.}
\end{figure*}

Naturally, one may ask; what does really cause this increase of orbital chaoticity? And is this increase associated closely to the rounding of the shape of the halo at the central parts of the configuration? To answer these questions a number of elements have to be examined. Firstly, one can look at the Poincar\'e section of the orbits. Such a section, recorded when the orbits cross the $v_x=0$ plane, reveals that when the dissipational component is present, velocities increase close to the center (Fig.~\ref{fig:fig4.ps}). This is not a surprise and is clearly associated with close encounters of the orbits with the central baryonic mass; the particles approach close to the baryonic mass and get deflected.

\begin{figure*}
\psfig{figure=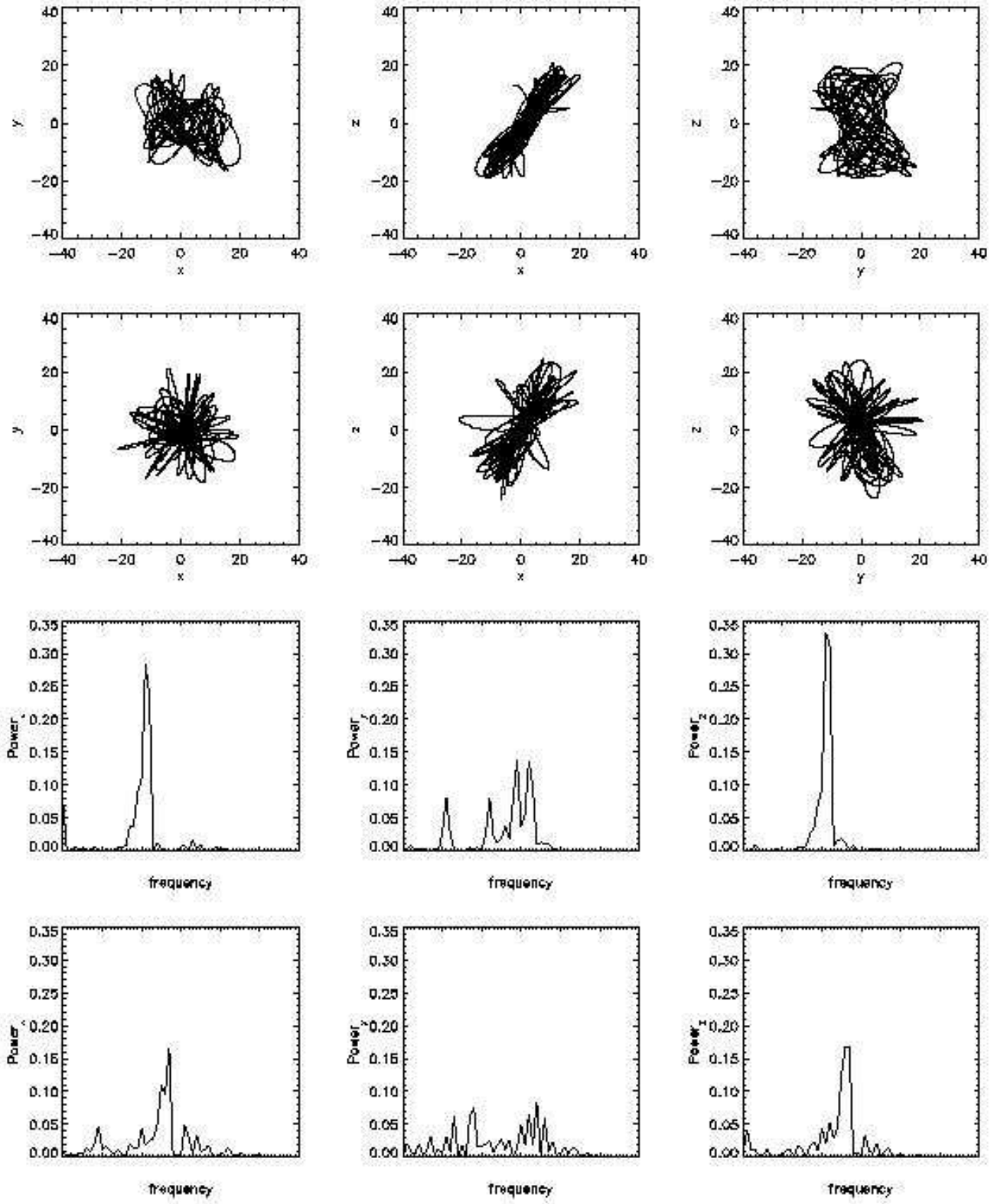,width=500pt}
\caption{\label{fig:fig5.ps}
First row panels:  x-y, x-z, and y-z projections of an orbit evolving in a halo without a central baryonic mass.
Second row panels: x-y, x-z, and y-z projections of the same orbit evolving in a halo with a central baryonic mass.
Third row panels: Fourier spectra of the first (without central mass) orbit.
Fourth row panels: Fourier spectra of the second (with central mass) orbit. The Fourier spectra in the three dimensions are broader in the second case. }
\end{figure*}

It is reasonable to assume that the close encounters with the central mass, and the time-dependence caused by the growth of the central mass can combine to cause a shift to the frequencies of motion of individual orbits with subsequent randomization of the direction of the orbital evolution. This way any regularity or sense of shape disappear, the orbit attempts to occupy phase space in many different directions, thus becoming highly chaotic. This process should be manifested as a broadening of the power spectra of orbits (since new frequencies appear fast). A look at the three frequency spectra of same orbit moving in the two different configurations (with or without a growing central baryonic element) makes obvious that the spectrum is much broader in the second case (Fig.~\ref{fig:fig5.ps}). The existence of the broader spectra obviously correlates to the increase of chaoticity observed in Fig~\ref{fig:fig3.ps}.

\begin{figure*}
\psfig{figure=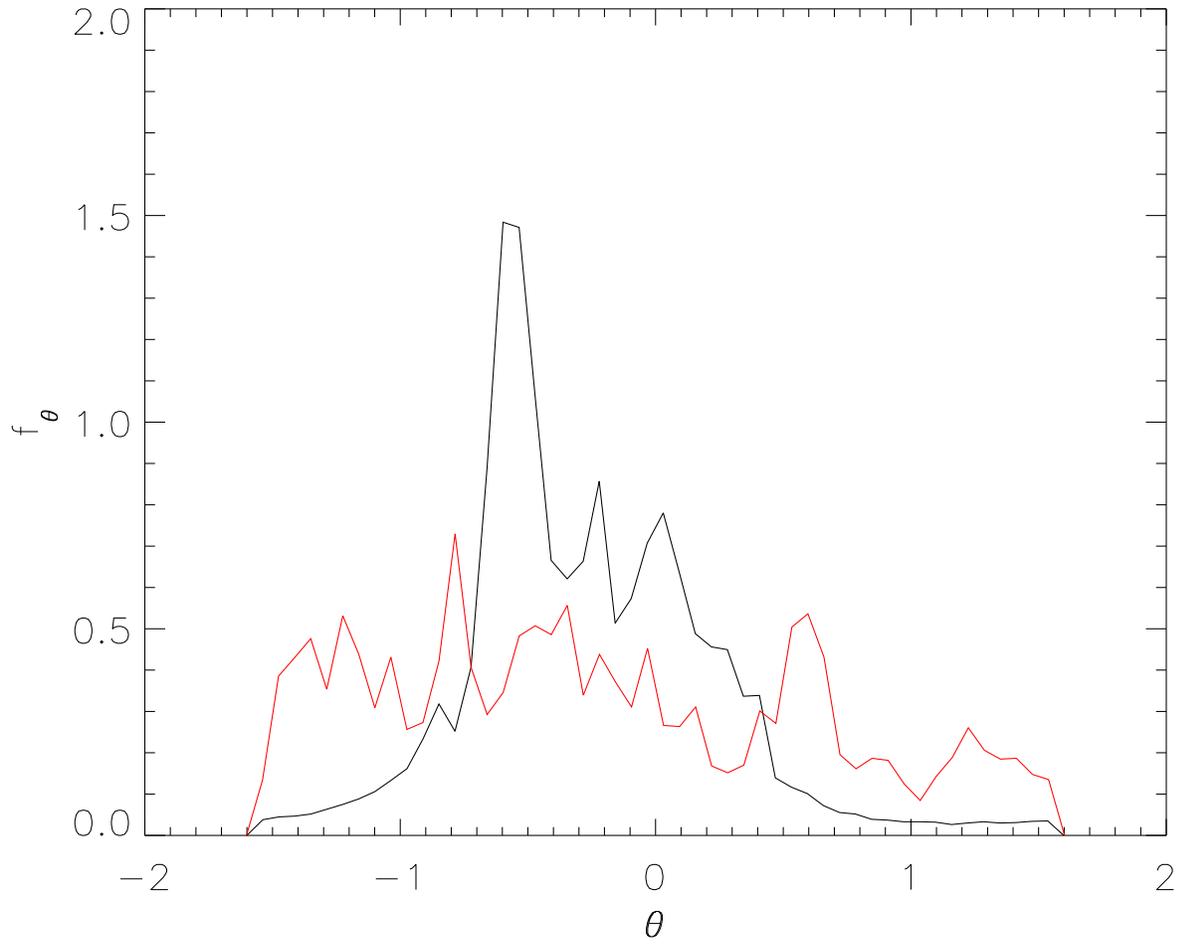,width=500pt}
\caption{\label{fig:fig6.ps}
Distribution of angles $\theta=tan^{-1}(x/y)$ explored for the dissipationless orbit (black) and the dissipational one (red). }
\end{figure*}

Although we have important hints about the dynamical process followed, one needs a careful quantification of if and how the direction of the emerging orbital components is really randomized. This effect can be visualized clearly for an orbit passing very close to the center. One may ask how the distribution of the angles $\theta=tan^{-1}(x/y)$ changes between the dissipational  and  dissipationless  regime. In  a  triaxial dissipationless  regime such  a distribution  should  have some preferences, because there are no intense phenomena like strong deflections. On the other hand a randomization of the orbit should be associated or approach a distribution closer to normal, since all possible angles must have equal probability to be visited. Indeed by overplotting the $\theta$ 
distributions for the same orbit in both regimes it is obvious that what was expected actually happens Fig~\ref{fig:fig6.ps}.

This randomization will not happen for every orbit in the system, but a significant
 percent of orbits  in our sample  followed the aforementioned process. By visual 
inspection of a number of orbits one can easily identify a significant percentage 
which try to mimic and are restrained in the box-orbits fashion. 
The close encounters with the central dissipational components will make these orbits 
rounder. It has to be emphasized that although box-like orbits with bigger amplitudes 
can in principle experience similar dynamics, one has to remember that as the amplitude
increases the randomization effect decreases simply because the number of orbital 
periods is smaller. This means that the opportunities for close encounters with the 
central gas are decreased. This seems to explain why the effect of rounding is 
limited only in the central parts. A second reason should be that the effect the 
central mass has on the rounding of the phase space weakens in bigger distances 
from the center. Box-like weakly chaotic orbits with big amplitude do not have 
enough time to ``break'' and explore many new directions during their evolution. 
In an imaginary system which evolves for far longer times we would probably 
observe a rounding of the density in bigger radii too. It is the evolution time of 
the system that puts a limit in the maximum radius it gets rounder because of 
deflections by a central growing baryonic mass.

%AND
Figure ~\ref{fig:fig1.ps} also shows that the shape modification is not constant 
with respect to time. It is faster in the first 5 Gyrs for both the axis ratios: 
$b/a$ and $c/a$. This is related by the total fraction of box-like orbits within the 
central region of the halo. 
The 3-dimensional distribution of the mass is intimately connected
to its orbital content. 
Jesseit et al. (2005) showed correlations between orbital fractions and the shape 
parameters as determined by the axis ratios of the
inertial tensor. To give a rough estimate on the box orbit fraction in our initial 
halo, we selected the merger remnant from the original sample of Naab \& Burkert (2003) 
which resembles the analysed DM halo the most. Hence we 
conclude that the DM Halo has a box orbit fraction of at least 0.67  for the inner 
10\% , 0.49 for the inner 20 \%, 0.39 for the inner 30\% , 0.33 for the inner 40\% , 
0.27 for the inner 50\% and 0.23 at 60\% of all particles.
The fast change in the shape in the first Gyrs is related to the high number of 
box-like orbits (i.e. highly modificable orbits) within the central region of the halo. 
%AND

\begin{figure*}
\psfig{figure=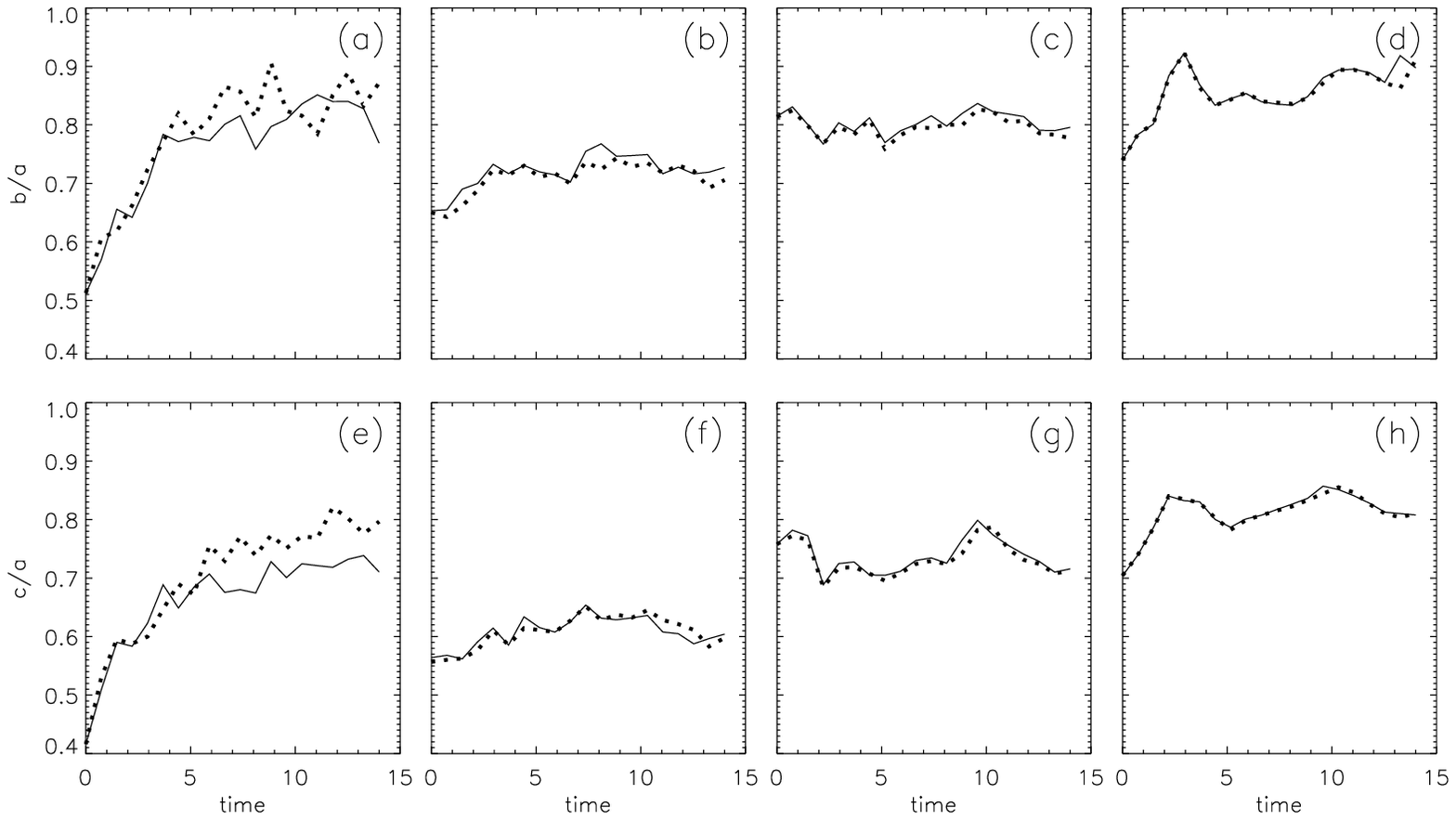,width=500pt}
\caption{\label{fig:fig7.ps}
Top panels: evolution of the axes ratio $b/a$ of two dissipational systems, one with $\epsilon=2.4$ (dotted line) and one with $\epsilon=0.6$ (solid line) for different distances from the center. 
(a) $r=14.9kpc$, 
(b) $r=163.4kpc$, 
(c) $r=311.9kpc$, 
(d) $r=460.4kpc$. 
Bottom panels: same as top but for axes ratio $c/a$. 
(e) $r=14.9kpc$, 
(f) $r=163.4kpc$, 
(g) $r=311.9kpc$, 
(h) $r=460.4kpc$.}
\end{figure*}

Some additional experiments were necessary in order to corroborate the understanding of the physics. First we performed the same simulation using  a much bigger  softening parameter  $\epsilon=2.4$. This simulation was necessary in order to show that the effects we observed where not caused by the numerics of the experiment. Comparison of the evolution of the axes ratios show clearly that the larger epsilon does not alter the results  (Fig.~\ref{fig:fig7.ps}). Moreover, the chaoticities of the orbits in this regime do not differ significantly from the ones evolving in a central growing mass regime (Fig.~\ref{fig:fig9.ps}). This is a strong indication that the observed effects are not numerical artifacts.

\begin{figure*}
\psfig{figure=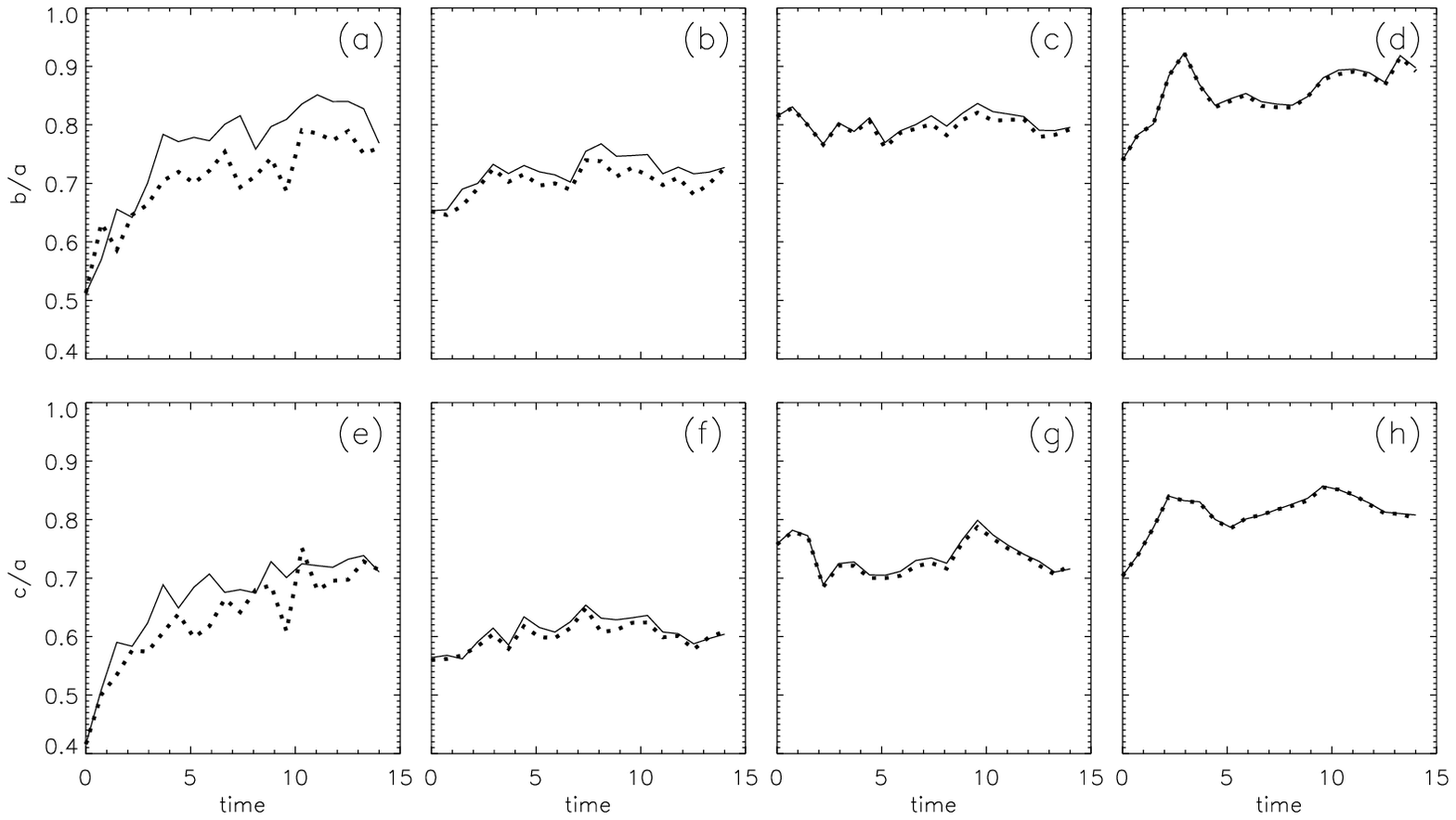,width=500pt}
\caption{\label{fig:fig8.ps}
Top panels: evolution of the axes ratio $b/a$ of two dissipational systems, one with 500 central masses (dotted line) and one with one central mass (solid line) for different distances from the center. 
(a) $r=14.9kpc$, 
(b) $r=163.4kpc$, 
(c) $r=311.9kpc$, 
(d) $r=460.4kpc$. 
Bottom panels: same as top but for axes ratio $c/a$. 
(e) $r=14.9kpc$, 
(f) $r=163.4kpc$, 
(g) $r=311.9kpc$, 
(h) $r=460.4kpc$.}
\end{figure*}

\begin{figure*}
\psfig{figure=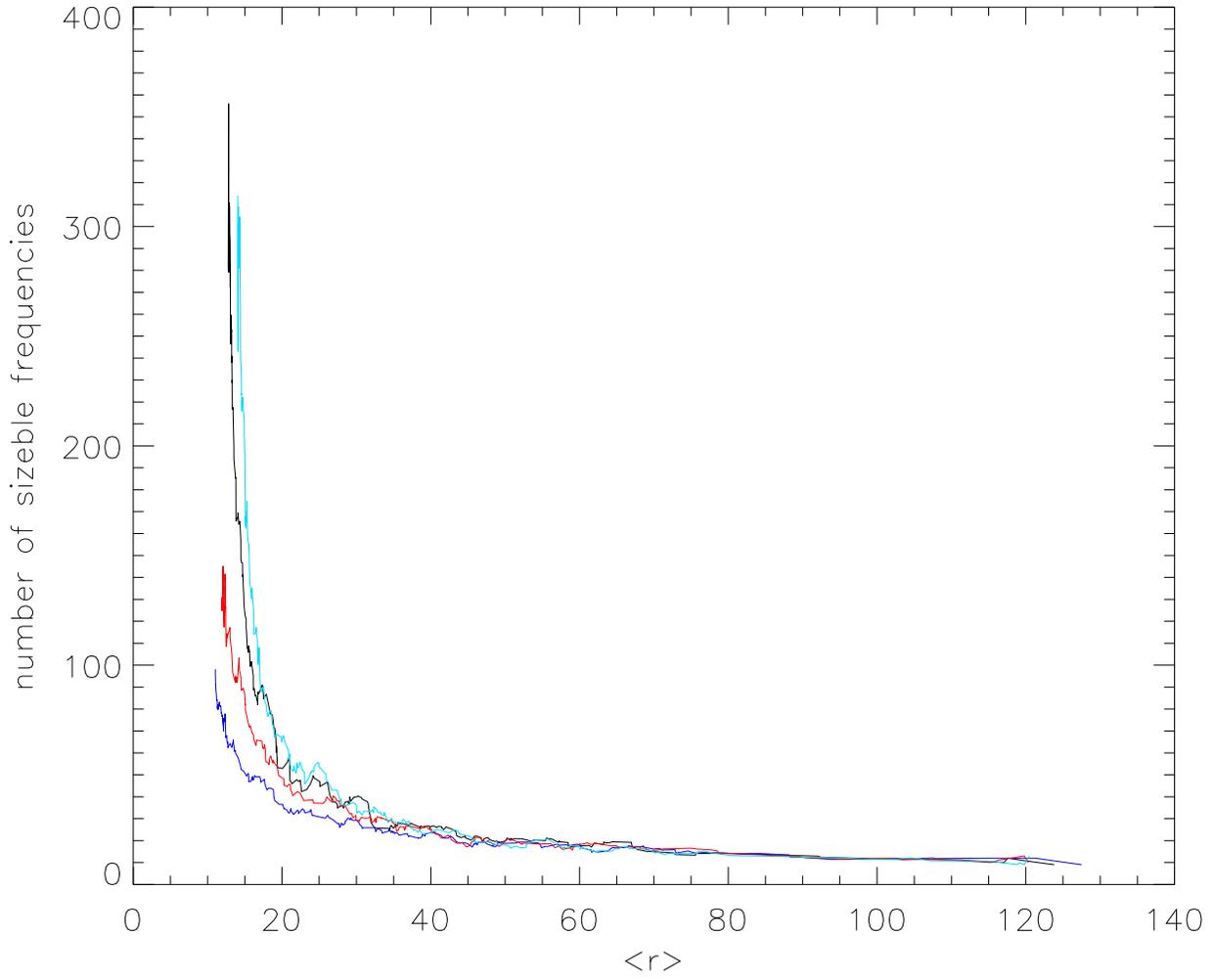,width=500pt}
\caption{\label{fig:fig9.ps}
Boxcar averaged complexities (number of important frequencies of motion) computed for 300 orbits versus mean radial distance from the center. 
Dark line: dissipational system with only one baryonic mass at the center ($\epsilon=0.6$). 
Red line:  dissipational system with 500 baryonic masses at the center ($\epsilon=0.6$). 
Blue line: dissipationless system ($\epsilon=0.6$). 
Cyan line: dissipational system with $\epsilon=2.4$. 
}
\end{figure*}

The second experiment was a more realistic version of the first one. Specifically, we replaced the central baryonic mass with a set of 500 smaller baryonic masses. The total mass and the rate of mass growth were kept the same. The 500 baryonic masses were localized at the center. By comparing the effects of the two dissipational simulations one can establish that the crucial aspects of dynamics are similar in both cases. Indeed, the evolution of the axes ratios are similar although they start  differing  as  we  approach very  close  to  the center (Fig~\ref{fig:fig8.ps}).  This is reasonable since the deflections are not very strong anymore, therefore the effect needs more time to become evident. An analysis of the chaos of the system suggests that the chaoticity of the 300 orbits we analyzed are in relatively lower level than when we had one central mass but still higher than the dissipationless case (Fig~\ref{fig:fig9.ps}).

\begin{figure*}
\psfig{figure=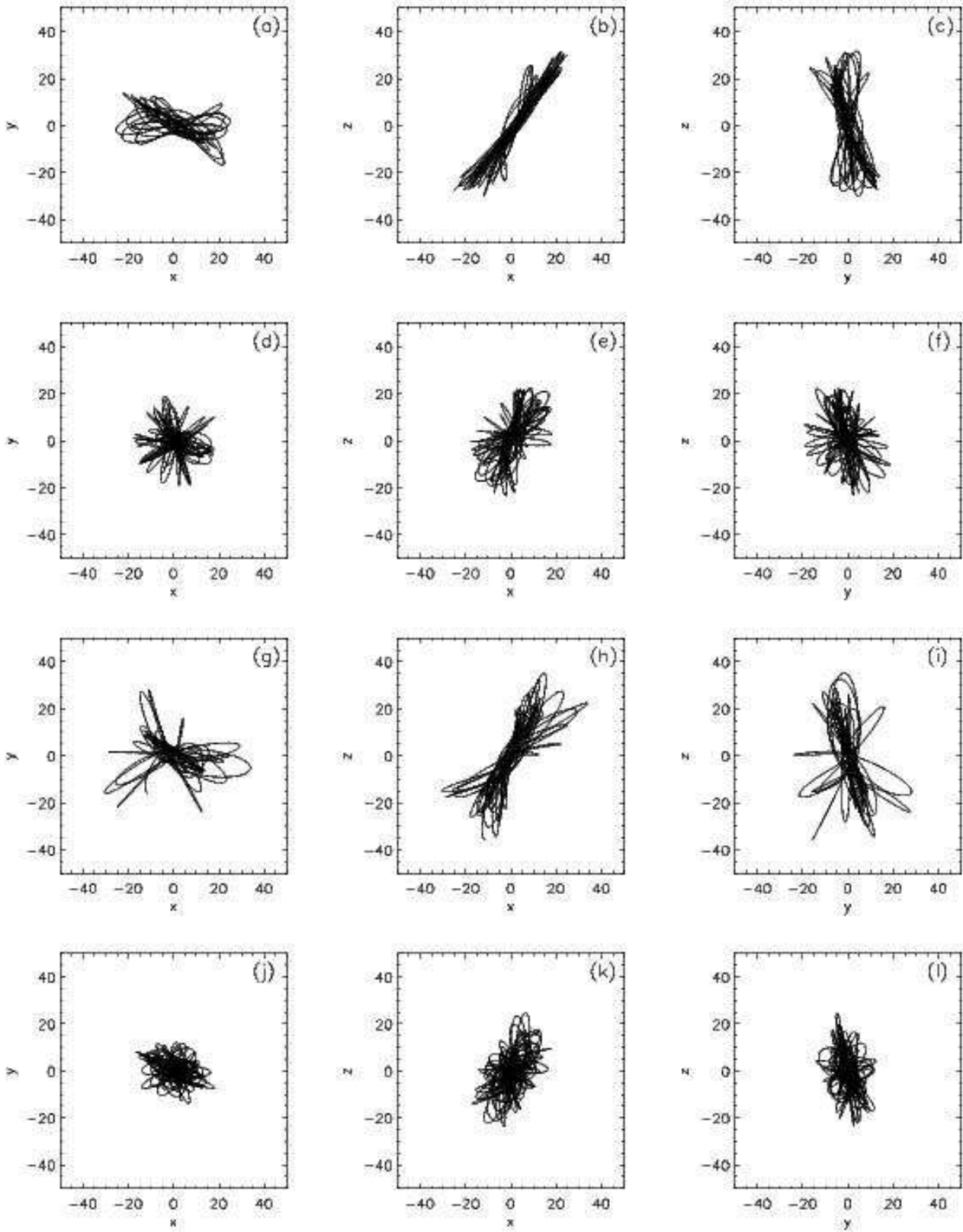,width=500pt}
\caption{\label{fig:fig10.ps}
Projections of the same orbit for 4 cases:
(a)-(c)dissipationless system ($\epsilon=0.6$).
(d)-(f)dissipational system with one central baryonic mass ($\epsilon=0.6$). 
(g)-(i)dissipational system with one central baryonic mass and softening $\epsilon=2.4$,
(j)-(l)dissipational system with 500 central baryonic masses ($\epsilon=0.6$).}
\end{figure*}

It is very instructive to see how the same orbit changes in the four different cases, (a) dissipationless system, (b) dissipational system, one central  mass, (c)  dissipational system one  central mass $\epsilon=2.4$, (d)dissipational system 500 central masses. Fig.~\ref{fig:fig10.ps} provides the three projections $x-y$, $y-z$, and $x-z$, of the same orbit for all the above cases. We chose this specific orbit because the difference between the dissipationless and dissipational cases are dramatic and thus make the phenomenon clearer. The effect we described is particularly obvious in the $y-z$ projection. In the dissipationless case the orbit has a strong radial component and passes very close to the center. When there is dissipation the central mass makes the potential rounder in the central areas where its influence is strong and it also deflects the orbit to several different directions. When $\epsilon=2.4$ the orbit also becomes thicker. For the case with 500 at the center the rounding effect is also present.

\section{Discussion and Conclusions}
\label{sec:discussion}
This investigation suggests that the dissipational component is responsible for repeatedly deflecting orbits which visit often the center of the system. Also the time-dependence associated with the growth of the central mass, (probably in combination with general nonlinearities of the triaxial configuration evolution), seem to shift the frequencies of the orbits,  making them extremely chaotic. These effects randomize the directions achieved during the evolution of the orbit. This randomization makes orbits explore many more angles, and therefore they occupy rounder shapes. When this effect takes place for a significant number of orbits it will be manifested in the density distribution as an approach to a rounder configuration.

This approach can easily serve as an explanation of why the rounding is limited only at the central parts of the configuration. Orbits with extensive amplitude cannot visit the center many times during the life-time of the system, thus they cannot be deflected repeatedly. This way their directions cannot be fully randomized in the sense this happens for orbits with smaller amplitude and they never get the opportunity to explore a much rounder space. Moreover, the influence of the central mass on the shape of the phase space decreases with increasing distance from the center. Since the shape of the orbits with bigger amplitude is not significantly altered by the existence of the central mass, there is no reason to expect any morphological alteration on the distribution in larger radii. 

It has to be noticed that this interpretation does not disagree with previous ones. It simply incorporates into the picture the role of extra influences into the evolutionary history of orbits by repeated central deflections, as well as the time-dependence of  the central baryonic mass. Two main questions are to be explained in future work in more controlled experiments: (a) What is the influence of each one of the two components? What is the exact role of the time-dependence of the central mass, and how does it combine with (or affected by) the nonlinearities of the self-consistent evolution in a triaxial regime? (b) Is there a general pattern in the way the frequencies of motion emerge for individual orbits? If yes how does this connect with the average location of the orbit as well as the rate of growth of the central mass? The answers to these questions can probably  decipher the fundamental dynamics of this phenomenon and provide important clues for the treatment of the problem using analytics.

\section*{Acknowledgments}

All the numerical simulations were performed on the zBox2 supercomputer
(http://www-theorie.physik.unizh.ch/$\sim$dpotter/zbox2/) at the University of
Z\"urich. AVM thanks Doug Potter for his fundamental help with all the zBox2 
issues.

\label{lastpage}

\end{document}